\documentstyle[prd,aps,epsfig]{revtex}
\begin{document} 
\def\beqra{\begin{eqnarray}}
\def\eeqra{\end{eqnarray}}
\def\beq{\begin{equation}}               
\def\eeq{\end{equation}}
\def\ds{\displaystyle}
\def\ts{\textstyle}
\def\ss{\scriptstyle}
\def\sss{\scriptscriptstyle}
\def\Vb{\bar{V}}
\def\phb{\bar{\phi}}
     \def\rhb{\bar{\rho}}
    \def\L{\Lambda}
\def\re#1{(\ref{#1})}
        \def\D{\Delta}
       \def\G{\Gamma}
\def\p{\partial}
 \def\de{\delta}
\def                     \lta                     {\mathrel{\vcenter
     {\hbox{$<$}\nointerlineskip\hbox{$\sim$}}}}
      \def      \gta
     {\mathrel{\vcenter
     {\hbox{$>$}\nointerlineskip\hbox{$\sim$}}}}
                   
     \renewcommand{\Re}{\mathop{\mathrm{Re}}}{\renewcommand{\Im}
     {\mathop{\mathrm{Im}}}
 \newcommand{\tr}{\mathop{\mathrm{tr}}}
     \newcommand{\Tr}{\mathop{\mathrm{Tr}}}
 
%
%
\def\i{i}
 \def\f{f}
 \def\d{d}
 \def\e{e}
 \def\half{\mbox{\small
$\frac{1}{2}$}}  
\sloppy  
\input  epsf.sty     
\twocolumn
\draft
\title{{\Large  {\bf  Boltzmann Suppression of Interacting  Heavy Particles}}}
\author{{\large  Patrizia
Bucci$^1$ and Massimo Pietroni$^2$}}
 \address{{\it $^1$ Dipartimento
di  Fisica, Universit\`a  di Padova,}\\
  Via F.  Marzolo  8, I-35131
Padova, Italy\\
 $^2$ INFN - Sezione di Padova,\\
 Via F. Marzolo 8,
I-35131   Padova,  Italy}
   \date{September  7,   2000}
  \maketitle
\abstract{Matsumoto and Yoshimura have recently argued that the number density
of heavy particles in a thermal bath is not necessarily Boltzmann-suppressed
for $T\ll M$, as power law corrections may emerge at higher orders in 
perturbation theory. This fact might have important implications on the
determination of WIMP relic densities.
On the other hand, the definition of number densities in a interacting 
theory is not a straightforward procedure.  
It usually requires renormalization of composite operators
and  operator mixing, which obscure the physical interpretation
of the computed thermal average.
We propose a new definition for the thermal average of a composite operator, 
which does not require any new renormalization counterterm and is thus
free from such ambiguities. Applying this definition to the annihilation
model of
Matsumoto and Yoshimura we find that it gives 
number densities which are Boltzmann-suppressed at any order in perturbation 
theory. We discuss also heavy particles which are unstable
 already at $T=0$, showing that power law corrections do in general emerge 
in this case. }
\begin{flushright}
DFPD 00/TH/41
\end{flushright}
\section{Introduction} 
 The computation  of number densities  of cosmological relics,  such as
weakly
interacting  
massive particles  (WIMPs), neutrinos, baryon
and lepton number,  is usually
accomplished 
by  using a thermally
averaged  Boltmzann equation \cite{LW,KT}.  The picture emerging in
this context is that of  a sudden freeze-out:  the particle number
density follows  the equilibrium value $N(T)$  for temperatures higher
than the freeze-out value,  $T_{f}$, below which the annihilation is
frozen.   In  the  WIMP  scenario  for  dark  matter,  the  freeze-out
temperature turns out  to be typically much lower  than the WIMP mass,
then one expects a  Boltzmann-suppression factor $\exp(-M/T_f)$ in the
final  WIMP  number  density. In  a  series  of  papers  \cite{MY},
Matsumoto and  Yoshimura (MY)  have challenged the  above conventional
conclusion, claiming  that two-loop corrections to  the number density
exhibit only  power-law suppression  in $T_f/M$, thus  dominating over
the  Boltzmann-suppressed  tree-level  contribution  if $T_f$  is  low
enough.  If confirmed,  such a  finding would  imply that  the present
constraints  on dark  matter  models are  actually underestimated  and
should be carefully  reconsidered. To be definite,  MY introduced a
toy model of two real scalar fields with lagrangian 
 
\begin{eqnarray}
{\cal L} &  = & \half \partial_\mu \phi \partial^\mu  \phi - \half M^2
\phi^2 -
  \frac{\lambda_{\phi}}{4!} \phi^4 +\half  \partial_\mu \chi
\partial^\mu  \chi  -
  \half  m^2   \chi^2  \nonumber  \\
  &  &  -
\frac{\lambda_{\chi}}{4!}       \chi^4-\frac{\lambda}{4}      \phi^2
\chi^2\,,\label{model}
\end{eqnarray}

  with $M  \gg m$.  The following relations  were assumed among the
coupling           constants\[
           |\lambda_\phi|          \ll
\lambda^2\,,\;\;\;\;\;\;|\lambda|\ll|\lambda_\chi| < 1\,,
\]
so that the  light $\chi$ particles act as an  efficient heat bath for
the heavy $\phi$'s. They  then considered the quadratic  part of
the Hamiltonian for $\phi$,  
 \beq 
 H_\phi = \half \dot{\phi}^2
+ 
  \half (\nabla  \phi)^2 +\half M^2  \phi^2\,,
 \eeq 
  and, for
temperatures $T$ such  that $m \ll T \ll M$  they defined the number
density of  $\phi$ particles as 
  
\beq
 N_\phi=\frac{\langle H_\phi
+{\mathrm counterterms}\rangle_T}{M}\,\
\label{Ynumber}
\eeq 
where MY define the  thermal average for an operator $A$ as 
\beq \langle A \rangle_T \equiv  \frac{\tr A e^{- H/T}}{\tr e^{- H/T}}
- 
                \frac{\langle                0|A|0\rangle}{\langle
0|0\rangle},\label{Yaverage}
    \eeq
$H$    being    the   total Hamiltonian.   At   tree-level,   the   well  
known   Boltzmann-suppressed
contribution is obtained
\[ N_\phi^{(0)} = \left(\frac{M T}{2 \pi}\right)^{3/2} \e^{-M/T}\,.
\]
The power-law contribution arises at  two-loops and is given by \beq
N_\phi^{(2)}=                        c                       \lambda^2
\frac{T^6}{M^3},\,\,\,\,\,\,\,\,\,c=\frac{1}{69120}\,,
 \eeq
 which,
as  anticipated, dominates  over  the tree-level  contribution at  low
temperatures. The interpretation of (\ref{Ynumber}) as the physical
number density  of heavy $\phi$ particles has  been later
questioned by  Singh and Srednicki  \cite{SS,S}. First,  they pointed
out that  the splitting of the  full Hamiltonian in two parts -- 
one  for the  ``system" of  $\phi$ particles,  and the  other  for the
``environment" of  $\chi$'s -- is problematic,  as the interaction
energy is not properly taken into account. It was actually shown
to be  of the same order of the  ``system" energy in  an exactly
solvable  Caldeira-Leggett model. Second,  Srednicki \cite{S}  considered a
modification of  the model in  eq.~(\ref{model}) by  taking complex
$\phi$ fields  charged under a  $U(1)$ symmetry. Then, he  showed that
from the fluctuations of the conserved charge it is possible to define
a number density which  exhibits the expected Boltzmann-suppression at
low temperature. Extrapolating this analysis to the MY model, where no
conserved charge exists, he proposed that a physical definition of the
number  density  could  be   given  as  $N_S=(1/M)  \rho_{BS}$,  where
$\rho_{BS}$ represents  those terms in the {\it  total} energy density
which have  a Boltzmann-suppression factor.  Moreover,  he argued that
non-Boltzmann-suppressed terms should be interpreted as corrections to
the energy density of the  $\chi$'s obtained after integrating out the
heavy $\phi$'s  at low temperature. Conclusive insight on  the last
point was  achieved by  Braaten and Jia  \cite{BJ}, who worked  in the
effective lagrangian approach. They were  able to show that the energy
density of  the $\chi$-fields actually  receives power-law corrections
from the exchange of virtual  $\phi$'s. However, the leading terms are
only  $O(\lambda^2  T^8/M^4)$. A  $O(\lambda^2  T^6/M^2)$ term,  which
would  correspond  to  the effect  claimed  by  MY,  was shown  to  be
eliminated by a field redefinition, thus unphysical. From the above
discussion, we  see that the problem  of the definition  of the number
density for a heavy particle in a {\it interacting} theory is still an
open  issue.  As  discussed  in  \cite{BJ}  the  effective  lagrangian
approach cannot  be of help  in this respect,  as it accounts  for the
effect of  the heavy $\phi$'s on  the light $\chi$'s,  whereas what we
need is a  way to compute the effect of the  $\chi$'s on the $\phi$'s,
just the opposite. 

In this paper, we will show that it is possible to give a  
non-ambiguous definition of the number density of an interacting particle
at {\it any} temperature
and at any  order in perturbation theory. In the case  of the MY model
of   eq.~(\ref{model}),  our   definition   exhibits  the   expected
Boltzmann-suppression at low temperature. 
We will find it very convenient  to work in the  real-time
formalism of finite temperature field theory (RTF) \cite{LvW}, which 
has the advantage of keeping $T=0$ and finite temperature effects
well separated. Indeed, as we will discuss, the ambiguities
 in the conventional 
definitions of the number density are mostly due to the 
$T=0$-renormalization procedure. Being able to single out  
$T=0$  ultraviolet  (UV) 
divergences turns then out to be a great advantage when one is interested in 
genuine thermal effects.
Moreover, we
will  eventually be  interested in  using the
 defined  number density  in an
  out-of-equilibrium context,  and in
 this  respect the  RTF,  due  to its  close
  relationship with  the
 Schwinger-Keldysh  formalism,  is clearly  a  much better
  starting
 point than  the imaginary  time formalism.

The tree-level  propagator for a real scalar particle in the RTF 
has the structure
 \beq
 {\bf D_0}
= {\bf  D_0^{T=0}} +  {\bf D_0^T}  = i{\bf P[}\Delta_0{\bf  ]} +  i 
(\Delta_0-\Delta_0^*) {\cal N}(|q_0|) {\bf B}\,,
\label{RTFprop}
\eeq
where $\Delta_0^{(*)} =  (q^2-M^2 + (-) i \varepsilon)^{-1}$ is
the   tree-level  Feynman  propagator,    ${\cal   N}$   the Bose-Einstein
(Fermi-Dirac)     distribution    function,     ${\cal    N}(x)=
(\exp(x/T)-(+)1)^{-1}$    and   the   two
matrices ${\bf P}$ and ${\bf B}$ are defined as
\begin{eqnarray}
{\bf P[}a(q){\bf ]} & = & \left[
\begin{array}{cc}
a(q) & (a-a^*)(q)   \theta(-q_0)\\
   &  \\   
(a-a^*)(q)   \theta(q_0) & -a^*(q)
\end{array}
\right]\,\,, \nonumber \\
 & & \nonumber \\
{\bf B} & = & \left[ 
\begin{array}{cc}
1&1\\
 1&1
\end{array}
\right] , \nonumber \\
\end{eqnarray}
with $\theta(x)$ the Heaviside's step function. 

 The  full propagator  has the  same structure  as the  tree-level one,
eq. (\ref{RTFprop}),  with $\Delta_0^{(*)}$ replaced  by its all-order
counterpart         \cite{LvW}
        \beq
        \Delta^{(*)}(q)
=\frac{1}{q^2-M^2-\Pi(q_0,|\vec{q}|)+(-)i\varepsilon}\,\,,
\label{fulldelta}
\eeq
$ \Pi(q_0,|\vec{q}|)$ being the (renormalized) full self-energy 
at finite  temperature.  In the following,  we will refer  to $i {\bf 
P}[\Delta_0]$ and  $i {\bf P}[\Delta]$  as the ``$T=0$" parts  of the 
tree-level  and full  propagators, and  to $  i (\Delta_0-\Delta_0^*)
 {\cal N}(|q_0|)  {\bf B}$ and  $ i (\Delta-\Delta^*)  {\cal N}(|q_0|)
 {\bf B}$  as the  ``thermal" parts.  It  is important to  notice that
 what  we call  the  ``$T=0$"  part of  the  full propagator  actually
 contains thermal  corrections in the  self-energy, so {\it it  is not
 the  full $T=0$ propagator}.

 Our  discussion and
 results may  be summarized  as follows. We  will start  recalling how
 composite  operator  renormalization  induces  a mixing  between  the
 would-be number (or energy-density) operator for the $\phi$-field and
 operators   made  up   of  $\chi$-fields   (the   ``counterterms"  in
 eq.~(\ref{Ynumber})), which in general make a definition of a ``pure"
 $\phi$-number  quite cumbersome. These  divergences, and  the mixing,
 arise   as  the   field   operators  are   evaluated   at  the   same
 point. Diagrammatically, they  arise when the ends of  the two $\phi$
 legs are joined,  {\it provided both legs are  $T=0$ propagators}. It
 is inherently a  $T=0$ effect, since if any of  the two external legs
 is thermal, no divergence arises thanks to the UV cut-off provided by
 the Bose-Einstein function.  Then,  a thermal averaging procedure can
 be naturally defined, at least for quadratic operators, which is free
 from composite operator UV divergences.  

  If we insist in defining
 the number density starting  from the energy density
 $\rho_{\mathrm
 tot}$, we are led towards  a different splitting of 
 $\rho_{\mathrm
 tot}$  with  respect  to  that  considered by  MY.  The  $\rho_\phi$,
 $\rho_\chi$, and  $\rho_{\mathrm int}$  components into which  we split
 $\rho_{\mathrm  tot}$,  are such  that: {\it  i)} each  term is
 closed under operator mixing, so that the splitting is independent on
 the  renormalization  scale;  {\it  ii)}  no  $O(\lambda^2  T^6/M^2)$
 correction to  any of  the pieces in  which $\rho_{\mathrm  tot}$ 
 is  split emerges;
 {\it  iii)} 
 the  $\phi$-energy density  is Boltzmann-suppressed.
 Finally, we  will point out that  a better definition  for the number
 density, also exhibiting Boltzmann-suppression  at $T<M$ but valid in
 any temperature range,  can be given by considering  the thermal part
 of  the  full  propagator in  RTF.  
  
  The  paper is  organized  as
 follows. In sect.~II the basic features of composite operator
renormalization are illustrated for the case of the $\phi^2$ operator in the
MY model. In sect.~III a new   definition of thermal average is given, 
which does not require composite operator renormalization. In sect.~IV this 
definition is applied to the energy-momentum tensor in the MY model, 
showing that it leads to a  splitting of the total energy density in
$\phi$, $\chi$, and ``interaction'' 
pieces which differs to that considered by MY
and is not plagued by the usual ambiguities induced by operator mixing.
In sect V a definition of the number density valid at any 
temperatures and extrapolable even to the massless limit is given. Moreover,
it is shown how an extension of the MY model in which the heavy fields are 
complex  and carry a conserved charge naturally leads to a 
Boltzmann-suppressed definition of the heavy particle number density.
In sect.~VI the case of a heavy particle which is unstable already at $T=0$ is
discussed, showing that power law corrections do indeed emerge in this case.
Finally, in sect.~VII we discuss the relation of our work to that of refs.
\cite{MY,SS,S,BJ} and give our conclusions.
 
\section{Renormalization of $\phi_0^2$}
Before considering  the complete energy-momentum  tensor, the analysis
of the
 composite operator  $\phi_0^2$ (where the $0$-label indicates
the bare  fields,
 masses and  coupling constants) will be  useful in
illustrating  the   main  points
  of   renormalization  and  thermal
averaging. We start by considering the
 $T=0$  case. Due to the
local product of  fields making up the operator,  new UV
 divergences
are  induced when  it is  inserted in  a Green  function,  which are
generally   not   canceled  by   the   lagrangian  counterterms.   The
renormalization of
 these  divergences then requires the introduction
of new
  counterterms. Renormalized  operators can be  defined, which
are generically
  expressible as linear combinations of  all the bare
operators of equal or lower
 canonical dimensionality (for a thorough
discussion of  composite operator
 renormalization,  see for instance
\cite{C,B}).  In  the  model  (\ref{model}) the
  renormalization  of
$\phi_0^2$  then  induces a  mixing  with  $\chi_0^2$,  and the
  two
renormalized operators, $[\phi^2]$ and  
 $[\chi^2]$ can be expressed
as
\beq
\left(
\begin{array}{l}
\ds[\phi^2] \\ 
\ds[\chi^2]
\end{array}
\right)  =\left(\begin{array}{cc}   1-\delta  Z_{\phi\phi}&\delta
Z_{\phi\chi}\\
 \delta Z_{\chi\phi}&1-\delta Z_{\chi\chi}
\end{array}
\right)
 \left(
\begin{array}{l}
\phi_0^2 \\ \chi_0^2
\end{array}
\right)\,.
  \eeq
  In  minimal-subtraction  scheme  (MS)  the  bare
parameters can 
 be expressed  in terms of the renormalized ones as
\beqra                            
 &&\lambda_{i,0}=\mu^{4-D}\lambda_i
\left(1+\sum_{k=1}^{\infty}\frac{a_k^i(\{\lambda_j\};D)}{(D-4)^k}\right)\,,\nonumber\\
&&M_0^2=M^2    \left(1     +\frac{\delta    M^2}{M^2}\right)=    M^2
\left(1+\frac{\mu^2}{M^2}\sum_{k=1}^{\infty}\frac{b_k^M(\{\lambda_j\};D)}{(D-4)^k}\right)\,,\nonumber\\
&&m_0^2=m^2    \left(1     +\frac{\delta    m^2}{m^2}\right)=    m^2
\left(1+\frac{\mu^2}{m^2}\sum_{k=1}^{\infty}\frac{b_k^m(\{\lambda_j\};D)}{(D-4)^k}\right)\,,
\label{MS}
\eeqra
  where $\{\lambda_i\}= \{\lambda_\phi,\lambda_\chi,\lambda\}$, $D$ is the 
space-time dimensionality,
and  the important
  point is  
 that  the $a_k^i$'s,  $b_k^M$'s and
$b_k^m$'s are mass, momentum,  and
 temperature independent. 
 Using
eqs. (\ref{MS}) it is then possible  to prove that, for any renormalized
green function 
 $G$ the following relation holds
$$
 M_0^2  \frac{\partial G}{\partial M_0^2}+
  m_0^2 \frac{\partial
G}{\partial  m_0^2} =  
 M^2  \frac{\partial G}{\partial  M^2}+
 m^2
\frac{\partial G}{\partial m^2}\,,
$$
 showing  that, being the  right hand side manifestly  finite, the
insertion of  the combination $M_0^2  \phi_0^2 + m_0^2  \chi_0^2$ does
not  induce  any UV  divergence.  The  $\delta Z_{\phi\phi}$,  $\delta
Z_{\chi\chi}$ renormalization constants can then be expressed in terms
of the mass counterterms as
 \beq
 \left\{
\begin{array}{c}
\ds \delta  Z_{\phi\phi}=\ds -\frac{\delta M^2}{M^2}  - \frac{m^2}{M^2}\delta
Z_{\chi\phi}\\ \ds   \delta   Z_{\chi\chi}=-\frac{\delta   m^2}{m^2}   -
\frac{M^2}{m^2}\delta Z_{\phi\chi}
\end{array}
\right.  \eeq  whereas   the  remaining  counterterms  
  $\delta
Z_{\phi\chi}$, $\delta Z_{\chi\phi}$,
 can be computed as 
\begin{eqnarray}
\delta Z_{\phi\chi}  & = &\ds  -\left.\frac{\langle \phi_0^2(x) \chi(y)
\chi(z)\rangle}{\langle                                     \chi_0^2(x)
\chi(y)\chi(z)\rangle}\right|_{\mathrm
     DIV}\,,\,\,\,\,\,\nonumber
\\
 & & \nonumber \\ 
 \delta Z_{\chi\phi}& = & \ds -\left.\frac{\langle
\chi_0^2(x)      \phi(y)      \phi(z)\rangle}{\langle      \phi_0^2(x)
\phi(y)\phi(z)\rangle}\right|_{\mathrm DIV}\,.\nonumber
\end{eqnarray}

\section{Thermal average without mixing}
Our  ultimate goal  is to  build a  definition of  number  density for
$\phi$ particles out of the bilinear operator $\phi^2$ (or $H_\phi$ as
in the MY  case). From the discussion in the  previous section, we see
that  a major  obstacle in  this respect  is the  feature  of operator
mixing,  since there  seems to  be  an inevitable  ``pollution" by  the
$\chi^2$ operator (implied by $\delta Z_{\phi\chi},\delta Z_{\chi\phi}\neq 0$),
  preventing us to cleanly separate  between the two
particle species. Moreover the $\mu$-dependence of 
 $\delta Z_{\phi\chi},\delta Z_{\chi\phi}$ makes any definition of number 
density renormalization-scale dependent.

A diagrammatic  analysis in the RTF can help
us in identifying the $T=0$ nature of such pollution, and to 
 give a
definition  of thermal expectation  value for  $\phi^2$ which  is free
from  it.
   
  Let us   consider  $ \tr   e^{-H/T}  [\phi^2]
/ \tr   e^{-H/T}$. 
At  tree-level in  the RTF it  is given by  the two  diagrams in
fig. \ref{fig1}a  
\begin{figure}[htb]
\leavevmode
\hspace{0.5cm} \epsfbox{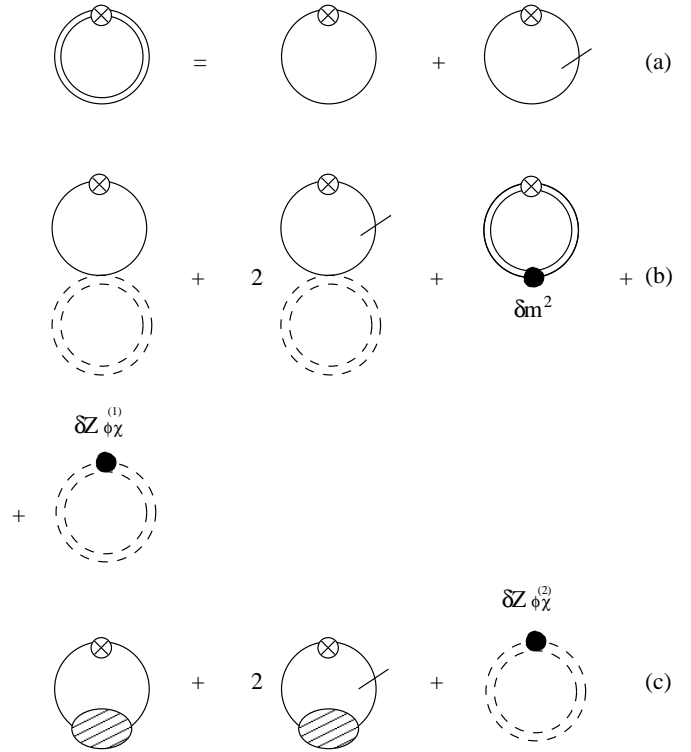}
\caption{Diagrammatic expansion of $\langle [\phi^2] 
\rangle_{T}$.}
\label{fig1}
\end{figure}
\noindent
where the  continuos line  represents  the $T=0$  part of  the
$\phi$ propagator, the barred continuos  line the thermal part and the
double line  the sum  of the two.  The first diagram  is quadratically
divergent and  requires a subtraction, whereas the  second one, thanks
to  the  Bose-Einstein  function  contained in  ${\bf  D_0^T}(q)$,  is
finite. At  $O(\lambda)$ the diagrams of fig.~\ref{fig1}b contribute.  
There are also   contributions   $O(\lambda_\phi)$   
(such   as   higher   order
contributions    containing   also    $\lambda_{\chi}$)    but   their
consideration is  not necessary for  our discussion, so we  will limit
ourselves to contributions containing only powers of $\lambda$. 
 The (quadratic) divergence due to the $T=0$ $\chi$-tadpole diagram (dashed
line)  is  cancelled  by  the  mass counterterm.  The  only  remaining
divergence comes from  the first diagram, in which both the
$\phi$  lines,  entering  the  cross  are  of $T=0$  type.  It  is  the
cancellation of this divergence which  calls into play the mixing, via
the  renormalization constant $\delta  Z_{\phi\chi}$ appearing  in the
last  diagram.   
At  $O(\lambda^2)$ (fig. \ref{fig1}c) the same  happens, where
\begin{figure}[h]
\leavevmode
\hspace{2.5cm} \epsfbox{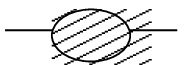}
\end{figure}
\noindent
is the self energy to $O(\lambda^2)$ renormalized by lagrangian 
counterterms.   
The  last
diagram, containing  a tree-level
 $\chi^2$ operator  multiplied by a
$O(\lambda^2)$  renormalization  constant is
  required  in order  to
cancel the divergence coming  from the integration over
 the momentum
of   $\phi$   particles   flowing    to   the   cross   with   $T=0$
propagators (the first diagram in fig. 1c).    
Note   that,   due    to   the    general   relation
$\sum_{i,j=1,2}\Sigma_{ij}=0$,  $\Sigma_{ij}$  being  the  self-energy  with
external
  thermal indices $i$  and $j$,  the contribution  with both
lines of thermal type
 also vanishes, once the summation over thermal
indices is  taken. The remaining
  loop divergences are  cancelled by
usual lagrangian counterterms and  therefore
 require no mixing.
 
The only divergence  which is not cancelled by  neither lagrangian nor
composite operators counterterms is the  tree-level one. Then, one usually
defines the  thermal expectation value of $\phi^2$  as the expectation
value of $[\phi^2]$  subtracted according to one of  the two following
ways  \cite{LvW,J},  
 \beqra
  \langle  [\phi^2] \rangle_T  &\equiv
&\frac{\tr [\phi^2]  
 e^{- H/T}}{\tr  e^{- H/T}} -  
 \frac{\langle
0|[\phi^2]|0\rangle}{\langle               0|0\rangle}              
\,\,\,\,\,\,\,\,\,\,\,\,\,\,\,{\mathrm     or}\nonumber\\
    &\equiv
&\frac{\tr [\phi^2] e^{- H/T}}{\tr e^{- H/T}} - \left.
 \frac{\langle
0|[\phi^2]|0\rangle}{\langle   0|0\rangle}
   \right|_{\mathrm   free
fields} \,\,\,\,\,\,\,\,\,\,\,\,\,\,\,.
\label{convthaverage}
\eeqra
 The definitions  above are both finite, and  thus a plausible
choice.  However,
 both of  them include  the contributions  from the
first and  the last graphs in 
figs.  \ref{fig1}b and \ref{fig1}c, that  is, they require mixing. 

The  above analysis  suggests a new  definition of thermal average  
of $\phi^2$,  which represents  the main  point of  this  paper, namely
\beqra
  \langle   \phi^2  \rangle_{\mathrm  Th}   &\equiv&  
  \int
\frac{d^4q}{(2\pi)^4}      D^T(q)\nonumber      \\     
      &=&\int
\frac{d^4q}{(2\pi)^4} 
 i(\Delta-\Delta^*)(q) {\cal N}(|q_0|)\,\,\,\,,
\label{thaverage}
\eeqra
 where  $ D^T(q)$  is any  of the four  component of  the {\it
thermal part of the full propagator}.
 
 Expanding $\Delta^{(*)}$ as
$\Delta^{(*)}=\Delta_0^{(*)}+\Delta_0^{(*)}   \Pi   \Delta_0^{(*)}   +
\dots$, and recalling the  relations between  $\Pi(q_0,|\vec{q}|)$ and
$\Sigma(q_0,|\vec{q}|)_{ij}$ \cite{LvW},
$$
  \Re  \Pi(q)  =  \Re \Sigma_{11}(q)\,,\,\,\,\,\,\,\,\,\Im  \Pi(q)  =
\varepsilon(q_0) \Im\left[\Sigma_{11}(q)+\Sigma_{12}(q)\right]\,,
$$
  one can  check that  the  perturbative expansion of fig.~\ref{fig1} is
reproduced, apart from the `problematic' diagrams, {\it i.e.}
the tree-level $T=0$ diagram,  the higher order
ones with  both $T=0$ lines flowing  into the cross,  and the diagrams
containing  composite operators counterterms,  which are  all omitted.
The definition  in (\ref{thaverage})  then automatically gets  rid of
all  the unpleasant  features of  
 the  conventional  definitions in
eq.   (\ref{convthaverage}),  namely,   the  need   of   an  arbitrary
subtraction and, more  importantly, composite operator renormalization
and mixing.  
 
  We are now  ready to  define a number  density for
$\phi$  which will not require any mixing with operators involving the
$\chi$-field and, as  a welcome
byproduct,  will  exhibit the  expected  Boltzmann-suppression at  low
temperature.
 
\section{Averaging the energy-momentum tensor}
The  renormalization of  the  energy-momentum tensor  is discussed  in
refs.\cite{LvW,B,J}. Expressing  it in  terms of bare  parameters, the
operator
$$                                                        
T^{\mu\nu}=
\partial^\mu\phi_0\partial^\nu\phi_0+\partial^\mu\chi_0
\partial^\nu\chi_0-g^{\mu\nu}
{\cal L}_0+\cdots\,,
$$
is finite when inserted in a Green function, that is, it does not require
extra counterterms besides those already present in the lagrangian.
The dots in the formula above 
  represent  pole   terms  proportional   to  $
(\partial^\mu \partial^\nu
 -  g^{\mu\nu} \partial^2)\phi_0^2$ and to
$ (\partial^\mu  \partial^\nu
 - g^{\mu\nu}  \partial^2)\chi_0^2$.
Since we are interested in the thermal average of $T^{\mu\nu}$
 alone,  translation invariance insures that such pole terms do not contribute.
On the other hand, a   subtraction    of    the   $T=0$    
contribution   as    in
eqs.  (\ref{convthaverage})  is   needed.  Working  with  renormalized
parameters,     $T^{\mu\nu}$      is     expressed     as
     \beq
T^{\mu\nu}=T^{\mu\nu}_{\mathrm can} +T^{\mu\nu}_{\mathrm c.t.}\,,
\label{tmunu} 
\eeq
 where $ T^{\mu\nu}_{\mathrm can}$  has the canonical form and $
T^{\mu\nu}_{\mathrm c.t.}$ contains  
{\it only the lagrangian counterterms}.
 
The finitness of $ T^{\mu\nu}$  means that the divergences induced by,
e.g.,  the  composite  operator  $\phi^2\chi^2$ are  canceled  by  the
lagrangian  counterterms and the  particular combination  of operators
$\phi^4$,  $\chi^4$, $(\partial \phi)^2,\ldots$, appearing  in $
T^{\mu\nu}$. In general, if we split $ T^{\mu\nu}$ in parts, each part
separately  will require  composite operator  renormalization. Indeed,
this is  what is done  by MY in  ref. \cite{MY}. They split  the total
hamiltonian as
$$
 T^{00}\equiv  H_{\mathrm tot} =  H_\phi+H_\chi+
 H_{\mathrm int}
+H_{\mathrm c.t.}\,,
$$
and compute thermal average of  $H_\phi$ (given in eq.~(2)).
In the RTF it gives
\begin{eqnarray}
\frac{\tr  H_\phi  e^{- H/T}}{\tr  e^{-  H/T}} &  =  &  
 \half  \int
\frac{d^4q}{(2\pi)^4}   \left[2q_0^2-    
   (q^2-M^2)\right]   \cdot
\nonumber \\  
 & & \nonumber \\
  & & \,\,\,\,\,\,\,\,\,\,\,\,\cdot
[{\bf D^{T=0}}+{\bf D^{T}}]_{12}(q)]\,,
\label{MYham}
\end{eqnarray}
where, differently from eq.  (\ref{thaverage}), also the $T=0$ part of
the full
 propagator appears. It  is important to recall here what we
have noticed after
 eq.(\ref{fulldelta}),  namely that the $T=0$ full
propagator  contains  thermal
  effects  via  the  self-energy.  Such
contributions  are not  cancelled  by purely
  $T=0$ subtractions  like
those defined in  eq.  (\ref{convthaverage}).  By  perturbatively
  expanding
$[{\bf D^{T=0}}]_{12}$  in (\ref{MYham})  at $O(\lambda^2)$  we find,
apart from  divergent contributions  which require  composite
operator
  counterterms  to be  added  to
$H_\phi$, the ``famous" 
 power-law contribution
 \beq
\begin{array}{l}
\ds \half  \int  \frac{d^4q}{(2\pi)^4}  
 \left[2q_0^2-  (q^2-M^2)\right]
\left.[{\bf   D^{T=0}}]_{12}(q)\right|_{O(\lambda^2)}   
    =   \\
\ds \half  \int  \frac{d^4q}{(2\pi)^4}  
 \left[2q_0^2-  (q^2-M^2)\right] 
\theta(-q_0) \\
\ds
\left[ i(\Delta_0^2-{\Delta_0^*}^2) \Re \Pi-
(\Delta_0^2+{\Delta_0^*}^2) \epsilon(q_0) \Im \Pi\right] 
  \,   \\
\ds= {\mathrm div.} + \frac{1}{69120}\frac{\lambda^2 T^6}{M^2}+\cdots\,\,,
\end{array}
\label{powerlaw}
\eeq
where $\epsilon(q_0)\equiv \theta(q_0)-\theta(-q_0)$.
The  same contribution, with  opposite sign, is found  from the
corresponding  piece  of  the   thermal  average  for  $H_\chi$.
  
Paralleling our  discussion in the  previous section, we  will instead
define   the  energy  density   of  the   $\phi$  field   as
  
\beq
\rho_\phi^\prime\equiv   \half   \int   \frac{d^4q}{(2\pi)^4}   \left[2q_0^2-
(q^2-M^2)\right]
  [{\bf D^{T}}]_{12}(q)\,,
\label{newdef}
\eeq
  and analogously
for $\rho_\chi^\prime$. 
The  complete energy density will then  be split as
\beqra
 \rho_{\mathrm tot}\equiv
  \langle H_{\mathrm tot} \rangle_T
&\equiv &
 \frac{\tr  H_{\mathrm tot} e^{- H/T}}{\tr e^{-  H/T}} - 
\frac{\langle      0|      H_{\mathrm     tot}      |0\rangle}{\langle
0|0\rangle}\nonumber\\
   &&=\rho_\phi^\prime   +\rho_\chi^\prime  
+\rho_{\mathrm int}^\prime\,,
\label{split}
\eeqra
so that the contribution of eq.~(\ref{powerlaw}) enters 
$\rho_{\mathrm int}^\prime$ instead of $\rho_\phi^\prime$.
Notice that  neither
$\rho_{\phi}^\prime$ nor  $\rho_{\chi}^\prime$ are affected by the $T=0$ 
subtraction.

The definition in (\ref{split}) exhibits three remarkable properties:

{\it i)} The splitting in (\ref{split}) is closed under operator 
mixing and composite operator renormalization. No composite operator 
counterterm is required to make $\rho_{\phi}^\prime$ or  $\rho_{\chi}^\prime$
separately finite and, since $\rho_{\mathrm tot}$ 
is also finite, the same is true 
also for $\rho_{\mathrm int}^\prime$.
Thus, our definition of number 
density, differently from that considered by MY, is independent on the 
renormalization scale $\mu$ \footnote{Of course, at any finite order in 
perturbation theory we still have the usual $\mu$ dependence induced by
lagrangian counterterms.}. There is no ``pollution'' from $\rho_{\chi}^\prime$
to $\rho_{\phi}^\prime$ at any value of $\mu$.

{\it ii)} None of the three pieces in (\ref{split}) contains 
$O(\lambda^2 T^6/M^2)$ terms. Indeed, the contribution of 
eq.~(\ref{powerlaw}) and  the analogous one for the $\chi$'s go both into 
$\rho_{int}^\prime$ and there, because of their opposite sign, cancel.

{\it iii)} $\rho_\phi^\prime$ is Boltzmann-suppressed for $T\ll M$.Indeed, writing $[{\bf D^{T}}]_{12}(q)$ explicitly
\beq 
[{\bf D^{T}}]_{12}(q) = -
\frac{ 2 \Im \Pi(q) \epsilon(q_0)}{(q^2-M^2-\Re \Pi(q))^2 + \Im \Pi(q)^2} 
{\cal N}(|q_0|)\,,
\label{bbss}
\eeq
we see that, due to the Bose-Einstein function,  
non-Boltzmann-suppressed contributions to the integral in eq.~(\ref{newdef})
might only come from momenta $|q_0| \lta T \;(\ll M)$. But, for these values of
momenta, it is the 
imaginary part of the self-energy which is Boltzmann-suppressed, as we show
explicitly in the Appendix. 

\section{A BETTER DEFINITION FOR $N_\phi$}
Using the $\phi$-energy density as given in eq.~(\ref{newdef}), 
we can now define the $\phi$ number density for $T \ll M$ in a way 
analogous to MY's definition, eq.~(\ref{Ynumber}), that is
\[ 
N_\phi = \frac{\rho_\phi^\prime}{M}\;\;\;\;{\mathrm for\;\;\; T\ll M}\;.
\]
However a better definition, valid at {\it any} value
of $T$, and again based on the average of a quadratic operator, can be 
given as
\beq
N_\phi \equiv \int \frac{d^4 q}{(2\pi)^4} |q_0| [{\bf D^{T}}]_{12}(q)\,.
\label{verynew}
\eeq
As $\rho_\phi^\prime$, eq.~(\ref{verynew}) exhibits the nice features of
 not requiring composite operator renormalization and of being 
Boltzmann-suppressed at low temperatures, but in addition it may be extended 
to 
high temperatures or, equivalently, to the massless limit.
The origin of eq.~(\ref{verynew}) can be understood considering the case of 
a complex scalar field, $\Phi$, in case an exact $U(1)$ symmetry is imposed to
the theory. In this case, a natural definition of number 
density involves the current operator
\[
j_\mu (x) \equiv \frac{i}{2} \left[ \Phi^\dagger (x) 
\stackrel{\leftrightarrow}{\partial_\mu} \Phi (x) - \Phi (x) 
\stackrel{\leftrightarrow}{\partial_\mu} \Phi^\dagger (x) 
\right]\;.
\]
In the RTF the thermal average of this operator corresponds to 
$i \partial_{x_\mu} \left.  D_C(x-y)\right|_{x=y}$ where 
$D_C\equiv [{\bf D}]_{12}+[{\bf D}]_{21}$.
The thermal part of the propagator for a complex scalar field differs from 
eq.~(\ref{RTFprop}) only in the replacement of ${\cal N}(|q_0|)$ with
$\tilde{\cal N}(q_0)\equiv \theta(q_0) {\cal N}_+(q_0)+
\theta(-q_0) {\cal N}_-(q_0)$, where ${\cal N}_\pm (x) = 
(\exp(\beta(x\mp \mu_\Phi)-1)^{-1}$ and $\mu_\Phi$ is the 
$\Phi$ chemical potential.
Since the current is conserved,
the evaluation at $x=y$ does not require composite operator renormalization. 
Indeed, it can be checked explicitly that, after Fourier transforming, we get
\beq
\langle j_0 (x) \rangle_T = 2 \int \frac{d^4 q}{(2\pi)^4} q_0 
[{\bf D^{T}}]_{12}(q) \,,
\label{corrcar}
\eeq
where the $T=0$ contribution vanishes. Incidentally, this  sheds 
new light on the conclusions of ref.~\cite{S} in which it was shown that
in case a $U(1)$ symmetry exists, a Boltzmann-suppressed number density
can be defined. From what we saw above, it is so because in this case the
current operator for $\Phi$ does not mix with any other operator. 
In absence of mixing, terms like (\ref{powerlaw}) are absent and the number 
density is automatically Boltzmann-suppressed.

 At tree-level we get the well-known result for the charge density
\[
Q_\Phi = \langle j_0 (x) \rangle_T = \int\frac{d^3 \vec{q}}{(2\pi)^3} 
\left[{\cal N}_+(\omega)-{\cal N}_-(\omega)\right]\;,
\]
where $\omega=(|\vec{q}|^2+M^2)^{1/2}$.

Now we can see how eq.~(\ref{verynew}) can be derived in analogy to 
(\ref{corrcar}).
The number density for the real scalar field $\phi$ should differ 
from the previous
quantity in two respects. First, particle and antiparticle contributions 
should be summed up, in order to account for their `Majorana' nature. This
is obtained by taking the modulus of $|q_0| $ inside (\ref{verynew}). 
Second, a factor $1/2$ is needed, in order to take into acccount  the 
number of degrees of freedom.

\section{The case of decaying heavy particles}

The discussion of the previous sections is straightforwardly generalizable to
any model in which heavy particles annihilate into lighter ones, both bosons
and fermions. The basic conclusions remain unalterated, the main reason 
being the 
Boltzmann-suppression of the imaginary part of the self-energy at small 
momenta (see the discussion at the end of Sect.IV and in the Appendix).

The situation changes drastically if the heavy particle is unstable  
at $T=0$. 
If, for definitness,  it decays into two lighter particles of mass 
$m < M/2$ the propagator exhibits an 
imaginary part for $2 m < p_0 < M$ already at
 $T=0$, so that there is no reason to expect that it is Boltzmann-suppressed 
for $T<M$. The heavy particle is actually a resonance, whose energy-momentum 
can take values much lower than the peak at $p^2=M^2$, of course paying the 
usual Breit-Wigner suppression far away from the peak. In this case, power-law
 contributions to the resonance number density do in general emerge, as we will
briefly discuss.

We will consider for definitness the annihilation model discussed in 
ref. \cite{decay} and in the first of refs. \cite{MY}, in which the interaction
  between the  heavy  $\Phi$ and the light $\chi$-bosons is due to the 
term
\beq
{\cal L}_{\mathrm int} = - \frac{\mu}{2}\Phi \chi^2\,,
\eeq
where we  assume  that $\mu \ll M$. In  this model, the 
imaginary part of the self-energy is $\propto \mu^2$, so that, using 
(\ref{bbss}) in (\ref{verynew}) we get 
\beq
N_\Phi \sim \frac{\Gamma_0}{M^3}T^5\,,
\eeq
with $\Gamma_0\equiv \mu^2/(32 \pi M)$ the $T=0$ decay rate on-shell.
In ref. \cite{decay} a different behavior, $N_\Phi \sim T^4$ was claimed.
Indeed, the following argument clarifies the physical origin of the power-law
contribution and confirms the $T^5$ behavior.

In the thermal bath, unstable $\Phi$ particles are continuously produced in 
$\chi$-$\chi$ annihilations. Being the $\Phi$ a resonance, the production 
energy, $\omega$, can be much smaller than $M$. 
The produced $\Phi$'s eventually decay with inverse lifetime 
$\Gamma=\mu^2/(32 \pi \omega) = \Gamma_0 M/\omega$.
Thus, the 
number density of $\Phi$'s of energy around $\omega$, $n^\Phi(\omega)$,
obeys the rate equation
\beq
\frac{d n^\Phi(\omega)}{dt}= \gamma - \Gamma n^\Phi(\omega)\,,
\eeq
where $\gamma$ is the rate per unit volume of the process
\beq
\chi\chi \rightarrow \Phi\rightarrow 
{\mathrm everything}\;.
\label{proc}
\eeq
The above equation leads to the equilibrium value
\[ 
n^\Phi_{\mathrm eq}(\omega) = \frac{\gamma}{\Gamma}\,.
\]
The production rate is given by $\gamma \sim N_\chi^2 \sigma v$, where the 
inclusive cross section $\sigma$  may be computed using the optical theorem as
$\sigma \sim \mu^2 \Im \Delta_\Phi / s$, $s$ being the square of the $\Phi$
four-momentum, and $\Delta_\Phi$ the full $\Phi$ propagator.

Taking $m\ll\sqrt{s} \sim T\ll M$, the initial $\chi$ particles are 
relativistic 
({\it i.e.} $N_\chi \sim T^3$) and we get $\sigma \sim \mu^4/(M^4 T^2)$.
Putting all together, we get a contribution to the total 
$\Phi$-number density from 
the region $\omega \sim T$ of order
\beq
n^\Phi_{\mathrm eq}(\omega\sim T)
 \sim \frac{\mu^2}{M^4}T^5 \sim \frac{\Gamma_0}{M^3} T^5\,.
\eeq
The $\Phi$'s with energies closer to the peak of the resonance 
live longer  but their contribution to the
number density  is 
Boltzmann-suppressed, as two highly energetic $\chi$'s are required in the 
initial state to produce them.

As the energy of the relevant $\Phi$'s is 
typically $O(T)$, the energy density is 
\[
\rho_\Phi \sim O\left(\frac{\Gamma_0}{M^3} T^6\right) + 
{\mathrm Boltzmann-suppressed}\,,
\]
and  $\rho_\Phi/\rho_\chi$ vanishes as $T^2$ at low temperatures.

Before concluding this section, we observe that the definition of a sensible
number density along the lines of eq.~(\ref{Ynumber}), {\it i.e.} starting 
from the free hamiltonian, does not seem quite appropriate in this case. 
Indeed, as we have just discussed, the width of the $\Phi$ plays a crucial 
role, allowing energy values much lower than the tree-level mass-shell. It 
seems that, if one wants to insist in looking for definitions like 
eq.~(\ref{Ynumber}), a free Hamiltonian for quasi-particles, along the lines 
discussed for instance in ref.~\cite{Weldon}, should rather be used.

\section{Conclusions}
The definition of particle number density in an interacting theory is a 
delicate matter. The necessity of giving a meaning to divergent composite 
operators calls into play operator mixing, so that a separation between 
different particle species turns out, in general, to be a renormalization 
scale dependent procedure. 

Our guideline in this paper was precisely to find a definition
of number density free from this problem. We discussed how composite
operator renormalization is inherently a zero temperature effect which can
be isolated from the physically relevant finite temperature effects. We showed
that in the RTF a proper  definition of the energy and number densities
comes out quite naturally and can be systematically computed at any order
in perturbation theory. 

In the framework of the MY annihilation model, 
our definition of thermal average for 
bilinear operators leads to a different splitting of the total energy density
with respect to that considered by MY.  Each piece is finite, {\it i.e.} it 
does not require extra counterterms with respect to those already present 
in the 
lagrangian. Moreover, as a welcome byproduct, we obtain a Boltzmann-suppressed
energy density for the heavy particle $\phi$. Finally, no power-suppressed 
terms of the type found by MY appear. In this respect, we agree  with 
refs.~\cite{SS,S,BJ}; the crucial point is how the interaction is treated. 
In our discussion, we concluded that diverging
contributions like eq.~(\ref{powerlaw}) should naturally go into 
the `interaction' part of the total energy, to make it finite. 
If this is done consistently,
$O(\lambda^2 T^6/M^2)$ terms cancel out in $\rho_{\mathrm int}^\prime$.
This is consistent with the conclusion by Braaten and Jia \cite{BJ} that these
terms are unphysical.

Our definition of the $\phi$-energy density is close in spirit to that proposed
by Srednicki in \cite{S}. However the detail is different. Namely, in 
Srednicki's definition, all the Boltzmann-suppressed terms should be included 
into the $\phi$ energy density, whereas we have such kind of terms also in 
$\rho_{\mathrm int}^\prime$, for instance the contribution depicted
in fig.\ref {4a}.
\begin{figure}[htb]
\leavevmode
\hspace{2.5cm} \epsfbox{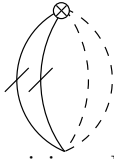}
\caption{A diagram giving a Boltzmann-suppressed contribution to 
$\rho_{\mathrm int}^\prime$}
\label{4a}
\end{figure}
\noindent
The interesting question, now, is how to obtain quantum kinetic equations
giving our number densities in the equilibrium limit. From the present 
 investigation it appears that the RTF may provide useful physical insights 
in this regard.

\acknowledgments
We acknowledge the financial support of the European Union under the 
contract HPRN-CT-2000-00149. The authors would like to thank also the Cern-TH
division, were this work was started.

\vspace{0.5cm}
\centerline{\bf APPENDIX}
\vspace{0.5cm}

The imaginary part of the $\phi$ self-energy in the MY model emerges at 
two-loops by the diagram in fig.~\ref{figA1}.
\begin{figure}[htb]
\leavevmode
\hspace{1.5cm} \epsfbox{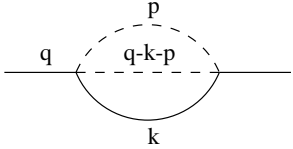}
\caption{Two-loop contribution to the self energy .}
\label{figA1}
\end{figure}
\noindent
From  eqs.~(\ref{newdef}), and (\ref{bbss}) 
we know that the contribution
to the integral defining $\rho_{\phi}^\prime$  from energies
 $|q_{0}|> T$ is Boltzmann-suppressed. 
Non Boltzmann-suppressed contributions might only come 
from $|q_{0}|\lta T(\ll M)$, but we will show at the end of this appendix 
that this is not the case as $\Im \Pi$ is Boltzmann-suppressed in this 
regime.  

According  to the standard  rules of
the RTF \cite{LvW} we  have for the
imaginary part:
$$
\begin{array}{rl}
\Im\Pi(q) = & -\half \epsilon (q_{0})(e^{\beta q_{0}}-1)\Sigma_{12}(q),
\end{array}
$$
 where
$$
\begin{array}{rl}
\Sigma_{12}(q) =  &\ds  -\frac{\lambda^{2}}{2}\int \frac{d^{4}p}{(2\pi)^4} 
\frac{d^{4}k}{(2\pi)^4}D_{12}(p)D_{12}(q-k-p)G_{12}(k),
\end{array}
$$
and
\beqra
iD_{12}(p)    &   =   &    2\pi   (\theta
(-p_{0})+{\cal N}(|p_{0}|))\delta(p^2)   \nonumber \\
iG_{12}(k)      &       =       &      2\pi       (\theta
(-k_{0})+{\cal N}(|k_{0}|))\delta(k^2-M^2)
\eeqra
the $\chi$  and $\phi$
particle  propagators  respectively.
Integrating  with respect to
$k_{0}$ and $p_{0}$ we obtain:
\begin{eqnarray}
\Im\Pi(q)           &          =          &           C          \int
\frac{d^3\vec{p}d^3\vec{k}}{\omega_{1}\omega_{2}\omega_{3}} 
  \{ \nonumber \\ 
&  &  \,\,\,\,\,\,             
[ \delta(q_{0}-(\omega_{1}+\omega_{2}+\omega_{3}))-\delta(q_{0}+
+\omega_{1}+\omega_{2}+\omega_{3}) ]   \cdot  \nonumber   \\  
   &  & \,\,\,\,\,\,     [(1+{\cal N}_{1})(1+{\cal N}_{2})
     (1+{\cal N}_{3})-{\cal N}_{1}{\cal N}_{2}{\cal N}_{3}]+
\nonumber      \\      
     &      &      \,\,\,\,\,\,     
[ \delta(q_{0}-\omega_{1}+\omega_{2}-\omega_{3})-\delta(q_{0}+
\omega_{1}-\omega_{2}+\omega_{3})] \cdot \nonumber \\ & & \,\,\,\,\,\,
 [(1+{\cal N}_{1}){\cal N}_{2}(1+{\cal N}_{3})-{\cal N}_{1}
 (1+{\cal N}_{2}){\cal N}_{3}]+ \nonumber \\ 
 & &                  \,\,\,\,\,\,                  
 [ \delta(q_{0}-\omega_{1}-\omega_{2}+\omega_{3})-\delta(q_{0}+
\omega_{1}+\omega_{2}-\omega_{3})] \cdot \nonumber \\ & & \,\,\,\,\,\,
[(1+{\cal N}_{1})(1+{\cal N}_{2}){\cal N}_{3}-{\cal N}_{1}
 {\cal N}_{2}(1+{\cal N}_{3})]+ \nonumber \\ 
 & &                  \,\,\,\,\,\,                  
[ \delta(q_{0}-\omega_{1}+\omega_{2}+\omega_{3})-\delta(q_{0}+
\omega_{1}-\omega_{2}-\omega_{3})] \cdot \nonumber \\ & & \,\,\,\,\,\,
[(1+{\cal N}_{1}){\cal N}_{2}{\cal N}_{3}-{\cal N}_{1}
 (1+{\cal N}_{2})(1+{\cal N}_{3})]\} 
\label{IM}
\end{eqnarray}
where   $C=\ds \frac{\lambda^{2}}{1024\pi^{5}}\epsilon(q_{0})$,   
$\omega_{1}=|\vec{q}-\vec{p}-\vec{k}|$,   $\omega_{2}=|\vec{p}|$   and  
$\omega_{3} = \sqrt{|\vec{k}|^2+M^2}$.
Since  $\Im\Pi$ is an
odd  function  of $q_{0}$  we  can restrict to 
$q_{0}>0$. $\Im\Pi$ can then be decomposed as:
$$
 \Im\Pi = \Im\Pi_{D}+\Im\Pi_{A}+\Im\Pi_{LD}
$$
  where:
\beqra
 \Im\Pi_{D}  &  =  &  C \int  \frac{d^3\vec{p}d^3\vec{k}}
{\omega_{1}\omega_{2}\omega_{3}}   \delta(q_{0}-
(\omega_{1}+\omega_{2}+\omega_{3}))     \nonumber     \\
     &     &  [(1+{\cal N}_{1})(1+{\cal N}_{2})(1+{\cal N}_{3})-
{\cal N}_{1}{\cal N}_{2}{\cal N}_{3}] \nonumber \\
\Im\Pi_{A} &  =   &      C   \int  \frac{d^3\vec{p}d^3\vec{k}}
{\omega_{1}\omega_{2}\omega_{3}}\delta(q_{0}-
\omega_{1}-\omega_{2}+\omega_{3})      \nonumber     \\
   &  & [(1+{\cal N}_{1})(1+{\cal N}_{2}){\cal N}_{3}-
{\cal N}_{1}{\cal N}_{2}(1+{\cal N}_{3})] \nonumber     \\
\Im\Pi_{LD}  &  = &  2C \int        
\frac{d^3\vec{p}d^3\vec{k}}{\omega_{1}\omega_{2}\omega_{3}}\delta(q_{0}-
\omega_{1}+\omega_{2}-\omega_{3})     \nonumber    \\    
 & & [(1+{\cal N}_{1}){\cal N}_{2}(1+{\cal N}_{3})-{\cal N}_{1}(1+{\cal N}_{2})
{\cal N}_{3}]  +  \nonumber  \\
 &  & -C\int   \frac{d^3\vec{p}d^3\vec{k}}{\omega_{1}\omega_{2}\omega_{3}}\delta(q_{0}+
\omega_{1}+\omega_{2}-\omega_{3})      \nonumber     \\
&  & [(1+{\cal N}_{1})(1+{\cal N}_{2}){\cal N}_{3}-{\cal N}_{1}{\cal N}_{2}
(1+{\cal N}_{3})]
 \nonumber \\ 
\label{A2}
\eeqra
 with ${\cal N}_{i}={\cal N}(\omega_{i})$.
  $\Im\Pi_{D}$ corresponds to the
decay  of an off-shell  $\phi$ into two
  on-shell  $\chi$'s
and  one on-shell $\phi $ (minus the inverse process) (fig.~\ref{figA2}a), 
$\Im \Pi_{A}$  to the
annihilation process, where an off-shell $\phi$  annihilates with an 
on-shell $\phi $  to produce two $\chi $'s 
(fig.~\ref{figA2}b), and $ \Im\Pi_{LD}$ 
 corresponds to Landau
damping of the $\phi$ induced by the scattering   with a   on-shell $\chi$
(fig.~ \ref{figA2}c). 
\begin{figure}[htb]
\leavevmode
\hspace{12cm}\vspace{0.5cm} \epsfbox{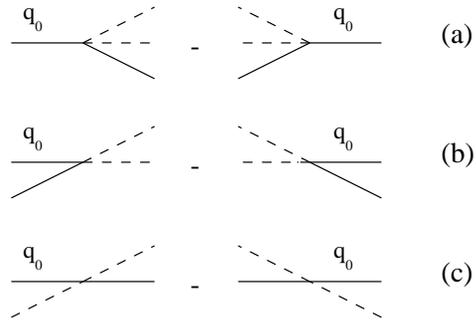}
\caption{a) Decay, b) Annihilation, c) Landau damping
.}
\label{figA2}
\end{figure}
\noindent
The computation  proceeds as in  \cite{HE} and,  for $\vec{q}
=0$, we get
\beqra
\Im\Pi_{D} & =  & C' \theta
(q_{0}-M)\int^{\bar{v}}_{a}dv
 f(v)f(\omega-v)\cdot \nonumber \\ 
 & &  \,\,\,\,\,\,\,\,\,   \ln\frac
{f(\omega^{2})f(\omega^{1}+v-\omega)}{f(\omega^{1})f(\omega^{2}+v-\omega)},
\nonumber  \\  
 \Im\Pi_{A}  &  =  & C'  [ \theta(q_{0}-M)\int^{+\infty }_{\bar{v}}dv   
+ \theta(M-q_{0})\int^{+\infty}_{a}dv]   \cdot\nonumber  \\
&  &  [(f(v)+1)f(\omega  +v)\ln  \frac{f(\omega^4)
f(\omega^3-v-\omega)}{f(\omega^3)f(\omega^4-v-\omega)}], \nonumber \\
\Im\Pi_{LD}  &  =  &
-C'\theta(M-q_{0})\int^{\bar{v}}_{a}dvf(v)(1+f(v-\omega)) \cdot\nonumber  \\
  &  & \ln \frac{(1+f(\omega^6))f(-\omega^5+v-\omega)}{(1+f(\omega^5))
f(-\omega^6+v-\omega)}+ \nonumber       \\
  &  & 2C'[\theta(q_{0}-M)\int^{+\infty}_{\bar{v}}dv                         +
\theta(M-q_{0})\int^{+\infty }_{a}dv]  \cdot  \nonumber  \\
 & & [f(v)f(\omega
-v)\ln\frac{(1+f(\omega^6))}{(1+f(\omega-v+\omega^6))}],
\label{FR}
\eeqra
 with 
$$
C'=\frac{\lambda^2}{128\pi^3}\epsilon(q_{0})(e^{\beta
q_{o}}-1)\frac{1}{\beta^2},
$$
and 
\beq
\begin{array}{lcr}
\ds a=\frac{M}{T},   & \ds\bar{v}=\frac{q_{o}^2+M^2}{2q_{0}T},  &
\ds v=\frac{\omega_{3}}{T},  \\
  &   &  \\
\ds \omega=\frac{q_{0}}{T},  &
\ds\omega^{1,2}=\half\frac{q_{0}-\omega_{3}\mp       k}{T}      &      
\ds \omega^{3,4}=\half  \frac{q_{0}+\omega_{3}\mp  k}{T},  \\
 &  &  \\
\ds \omega^{5,6}=\half \frac{\omega_{3}-q_{0}\mp k}{T}. & & 
\end{array}
\eeq
From the formulae above it is easy to check that 
the $\Im \Pi$ is Boltzmann-suppressed for $|q_{0}|\lta T(\ll M)$. 
Physically, it can be understood by
looking at fig.\ref{figA2}. For $|q_{0}|\lta T(\ll M)$ only the  annihilation 
 and the Landau damping  contribute. 
In the former case  (fig.~\ref{figA2}.b) 
the on-shell $\phi $ 
particle in the initial state comes from the heat-bath and then
carries a ${\cal N}$ factor. 
In the Landau damping case (fig.~\ref{figA2}.c), the energy to create
the on-shell $\phi$ has
to be provided by the $\chi$ from the
heat bath, so that a Boltzmann-suppressed ${\cal N}$ has  to be payed in 
this case too.

\end{document}